\documentclass[conference]{IEEEtran}
\IEEEoverridecommandlockouts
\usepackage{cite}
\usepackage{amsmath,amssymb,amsfonts}
\usepackage{graphicx}
\usepackage{textcomp}
\usepackage[dvipsnames]{xcolor}
\usepackage{subcaption}
\usepackage{booktabs}
\usepackage{textcomp}
\usepackage{multirow}
\usepackage{tikz}
\usepackage{pgfplots}
\usepackage{pifont}
\usepackage{algorithm}
\usepackage{makecell}
\usepackage[noend]{algpseudocode}
\usetikzlibrary{patterns}
\usepackage[a4paper, total={184mm,239mm}]{geometry}

\newcommand{\xmark}{\ding{55}}
\def\BibTeX{{\rm B\kern-.05em{\sc i\kern-.025em b}\kern-.08em
    T\kern-.1667em\lower.7ex\hbox{E}\kern-.125emX}}
\begin{document}

\title{AutoWS: Automate Weights Streaming in Layer-wise Pipelined DNN Accelerators}

\author{\IEEEauthorblockN{Zhewen Yu, Christos-Savvas Bouganis}
\textit{Imperial College London}\\
London, UK \\
\{zhewen.yu18, christos-savvas.bouganis\}@imperial.ac.uk}

\maketitle

\begin{abstract}

With the great success of Deep Neural Networks (DNN), the design of efficient hardware accelerators has triggered wide interest in the research community. Existing research explores two architectural strategies: sequential layer execution and layer-wise pipelining. While the former supports a wider range of models, the latter is favoured for its enhanced customization and efficiency. A challenge for the layer-wise pipelining architecture is its substantial demand for the on-chip memory for weights storage, impeding the deployment of large-scale networks on resource-constrained devices. This paper introduces AutoWS, a pioneering memory management methodology that exploits both on-chip and off-chip memory to optimize weight storage within a layer-wise pipelining architecture, taking advantage of its static schedule. Through a comprehensive investigation on both the hardware design and the Design Space Exploration, our methodology is fully automated and enables the deployment of large-scale DNN models on resource-constrained devices, which was not possible in existing works that target layer-wise pipelining architectures. AutoWS is open-source: https://github.com/Yu-Zhewen/AutoWS

\end{abstract}

\section{Introduction}
Recently, there is a broad interest in designing efficient FPGA accelerators for Deep Neural Networks (DNNs), aiming to optimize the trade-off between resource usage and performance. Two primary architectural strategies have emerged: sequential layer execution and layer-wise pipelining. The former involves the execution of DNN layers onto a single Compute Engine (CE) with time-multiplexing \cite{kathail2020xilinx}. In contrast, the latter strategy employs customized CE for each layer, interconnected in a chained manner for improved efficiency \cite{petrica2020memory}. 

A crucial difference between these two strategies is their memory requirements. Sequential layer execution stores both weight and activation data off-chip, leading to intensive off-chip memory access. Techniques such as tiling and double buffering are commonly employed to mitigate data communication bottlenecks by maximizing data reuse and hiding latency. Existing research on layer-wise pipelining \cite{venieris2017fpgaconvnet, petrica2020memory, fahim2021hls4ml}, on the other hand, restricts off-chip memory access to the inputs of the first CE and the outputs of the last CE, with intermediate activation data streamed through the CEs. Furthermore, all DNN parameters (weights) are preloaded into on-chip memory before inference begins. However, the substantial demand for the on-chip memory impedes the deployment of large-scale networks on resource-constrained devices.

\begin{figure}
    \centering
    \includegraphics[width=0.8\columnwidth]{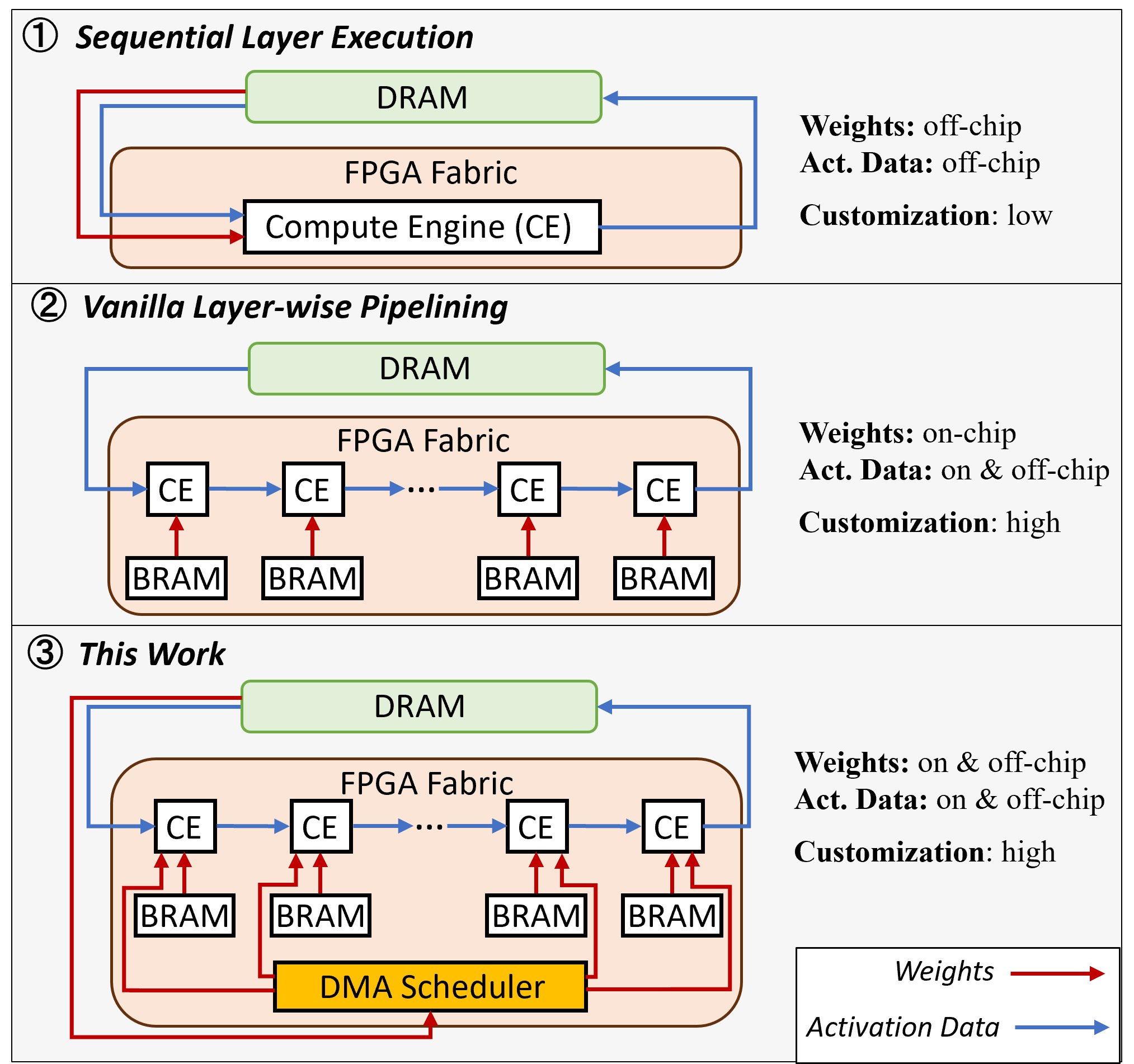}
    \caption{Comparison of existing designs and our architecture.}
    \label{fig:arch_compare}
    \vspace{-0.2cm}
\end{figure}

To overcome this bottleneck, this paper introduces a novel memory management methodology for the layer-wise pipelining architecture, that exploits both on-chip and off-chip memory for weights storage. Our approach introduces memory fragmentation and in this work we address the challenges of deciding the memory fragmentation parameters and bandwidth allocation when dealing with multiple pipelined CEs, offering an automated solution to these critical issues. Furthermore, we bundle the above memory subsystem with a parameterised layer-wise pipelining architecture resulting the first such accelerator that supports partial weight streaming. Figure~\ref{fig:arch_compare} illustrates the architectural distinctions between this work and existing solutions. Our contributions can be summarized as follows:

\begin{itemize}
    \item A scalable CE template featuring tunable unroll factors, and a flexible memory structure supporting static and dynamic weights loading, tunable in per-layer basis.
    \item A Design Space Exploration (DSE) process, which balances the processing rates of CEs and optimizes the allocation of off-chip bandwidth in an iterative and greedy way, based on our performance and resources modeling.
    \item A deterministic DMA scheduler that links off-chip memory to multiple CEs in a time-multiplexed manner, allowing the efficient exploitation of off-chip bandwidth with two clock domains.
\end{itemize}

\section{Related Works}

Early research on DNN acceleration focuses on sequential layer execution which only exploits intra-layer parallelization. Examples include DnnWeaver \cite{sharma2016high}, Angel-Eye \cite{guo2017angel} and Snowflake \cite{gokhale2017snowflake}. Subsequently, this line of research evolved into the use of the systolic array design, with investigations into efficient weight and activation data reuse strategies \cite{wei2017automated, samajdar2018scale}. Accelerators adopting the sequential layer execution architecture are often designed to be general-purpose. For instance, in Vitis AI \cite{kathail2020xilinx}, a single DPU IP configuration, leveraging a dedicated instruction set, can accelerate a wide range of different DNNs.

In contrast, layer-wise pipelining follows a distinct design methodology where the accelerator design is customized to each specific DNN workload, leading to improved performance. Previous work in layer-wise pipelining has predominantly focused on the efficient utilization of on-chip memory resources. Notable examples include fpgaConvNet\cite{venieris2017fpgaconvnet}, which relies on synthesis tools to determine the suitable resource type for weight storage (e.g., BRAMs or LUTRAMs), and hls4ml\cite{fahim2021hls4ml}, which provides users with control over this design aspect. DNNExplorer\cite{zhang2020dnnexplorer} incorporates this design choice into its Design Space Exploration (DSE) process. Furthermore, the authors of FINN \cite{petrica2020memory} observed that BRAMs might not be fully utilized due to parallel computation falling short of the fabric's provision. They addressed this issue by optimizing BRAM utilization through overclocking.

It is important to note that existing research on the layer-wise pipelining architecture has primarily concentrated on the optimization of on-chip memory only. In this paper, we demonstrate a novel memory management scheme that exploits both on-chip and off-chip memory, with the whole process automated. 

\section{Compute Engines}
Our architecture is depicted as \textcircled{3} in Figure~\ref{fig:arch_compare}. In this section, we focus on the internal structure of the proposed Compute Engine (CE), including its computational dataflow and memory structure, all of which can be tailored on a per-layer basis.

\begin{figure}
    \begin{minipage}{0.62\columnwidth}
        \includegraphics[width=\textwidth]{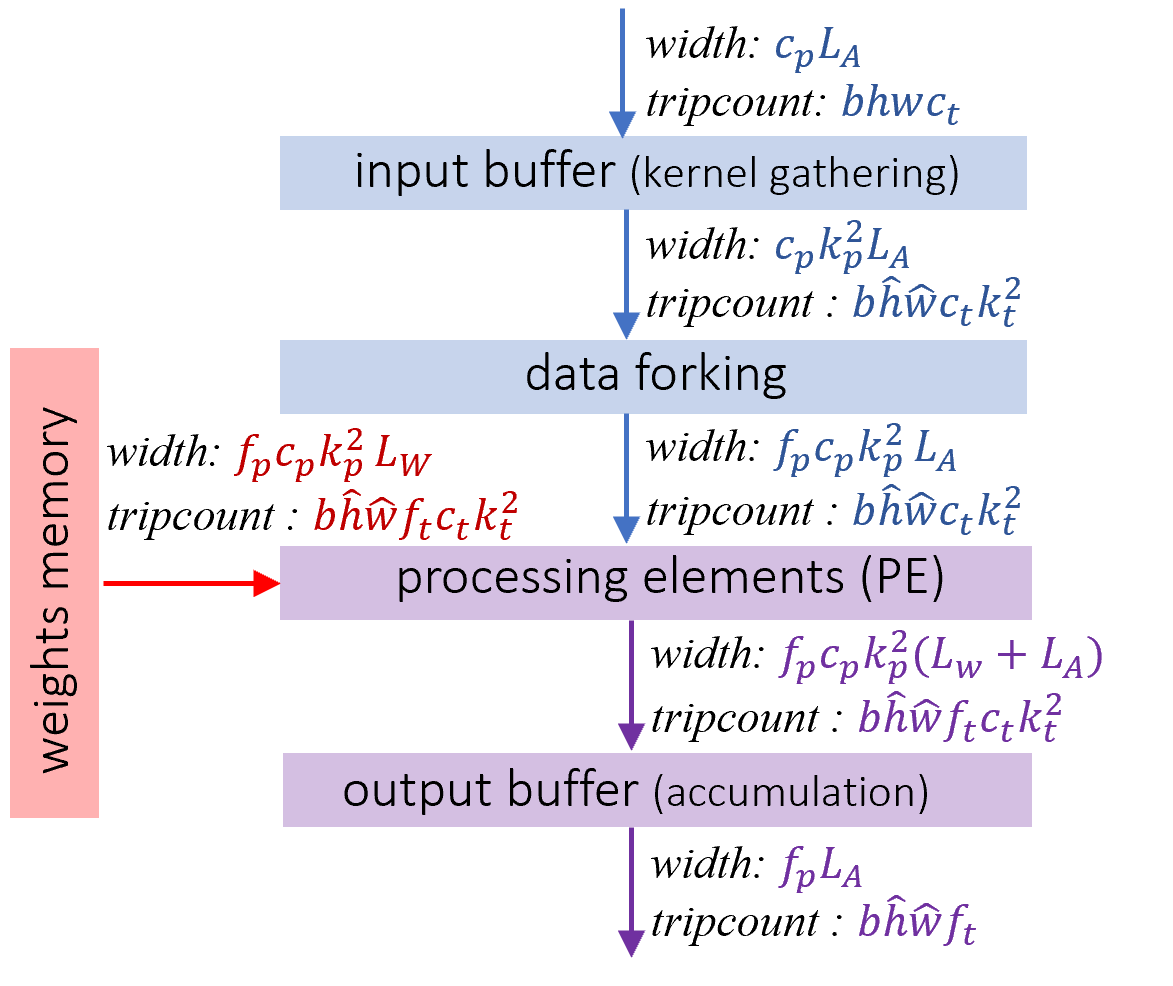}
    \end{minipage}
    \begin{minipage}{0.33\columnwidth}
        \begin{minipage}{\textwidth}
            \scriptsize
            \setlength{\tabcolsep}{2pt}
            \begin{tabular}{ll}
                \toprule
                \textbf{Symbols} & \textbf{Definitions} \\
                \midrule
                $b$ & batch size \\
                $c$ & input channel number \\
                $h, w$ & input height/width \\
                $k$ & kernel size \\
                $\hat{h}, \hat{w}$ & output height/width \\
                $f$ & filter number \\
                $L_{W}$ & weights bitwidth \\
                $L_{A}$ & activations bitwidth \\
                \bottomrule
            \end{tabular}
        \end{minipage}
        
        \medskip
        
        \begin{minipage}{\textwidth}
            \scriptsize
            $x_p$ and $x_t$ refer to the parallelism (unroll factor) and the tripcount of the loop iterates over any given symbol $x$, where $x=x_px_t$
        \end{minipage}
    \end{minipage}
    \caption{Dataflow of the compute engine.}
    \label{fig:ce}
\end{figure}

\subsection{Dataflow and Parallelization}
As depicted in Figure~\ref{fig:ce}, the dataflow involves interconnected building blocks, facilitated by FIFOs with handshake interfaces:

\begin{itemize}
    \item \textbf{Input buffer}: exists in convolution and pooling operations. A $k\!\times\!k$ 2D window slides over the spatial dimensions $h, w$  of activation data, implemented using shift registers to maximize data reuse.
    \item \textbf{Data forking}: exists only in convolution operations and duplicates the incoming activation data for $f$ copies, corresponding to $f$ different filters. 
    \item \textbf{Weights memory}: stores the weights in convolution and fully connected operations. More details on its implementation are provided in Section~\ref{subsec:w_stroage} and Figure~\ref{fig:fragment}.
    \item \textbf{Processing elements (PEs)}: an array of parallel processing elements that handle elementwise operations such as multiplication, addition, and ReLU activations. In cases where weights memory is not involved, this array may consume multiple activation data streams.
    \item \textbf{Output buffer}: utilized in convolution, this buffer accumulates incoming activation data streams across the 2D window and channel dimensions.
\end{itemize}

\subsection{Weights Storage}
\label{subsec:w_stroage}
\begin{figure}
    \includegraphics[width=\columnwidth]{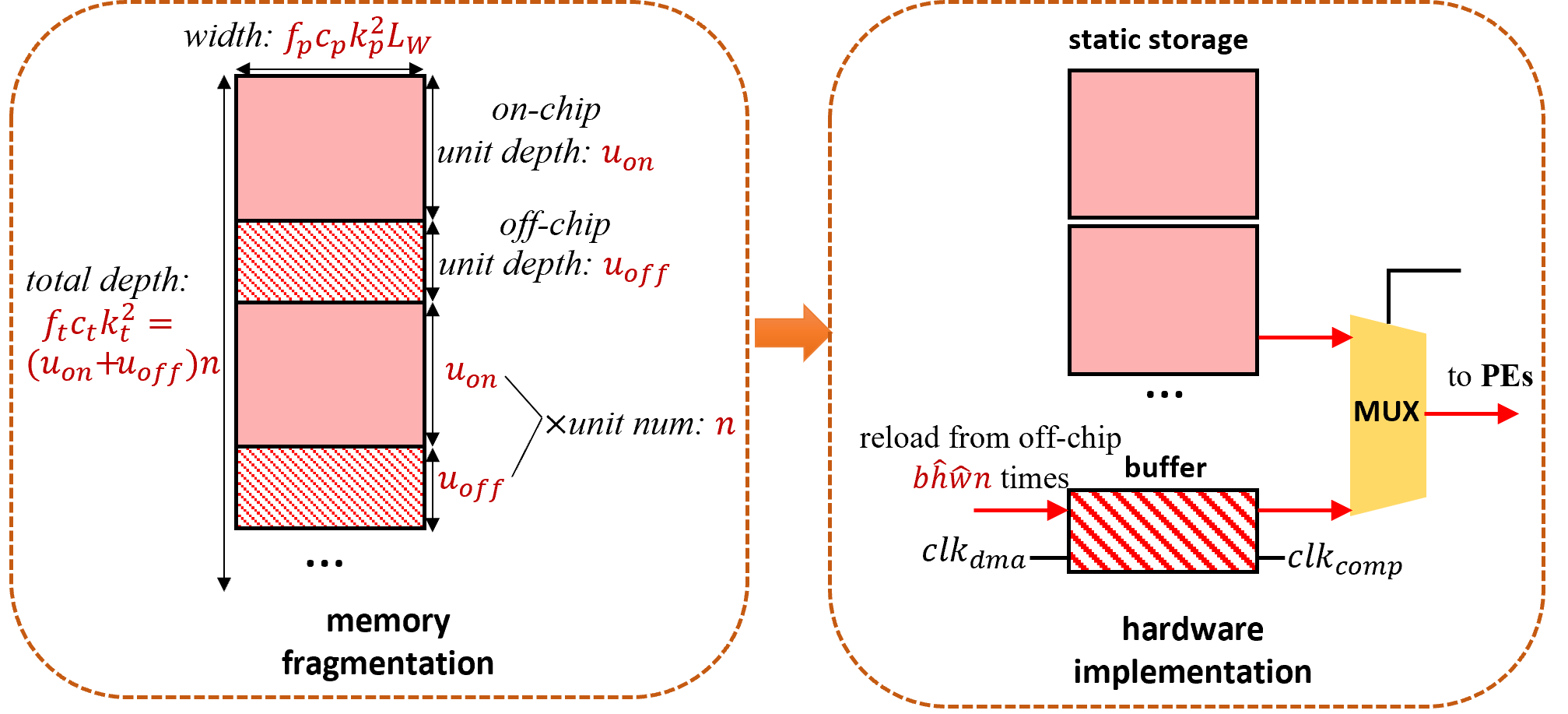}
    \caption{Fragmentation of the weights memory. The memory structure is split into the static regions that stay on-chip all the time and the dynamic regions which are reloaded from off-chip.}
    \label{fig:fragment}
\end{figure}

For easy illustration, we discuss the weight storage for convolutional layers, as any fully connected layer can be generalized to the case that $k$, $h$, $w$ are equal to one. In existing layer-wise pipelined designs\cite{venieris2017fpgaconvnet, petrica2020memory}, the required on-chip memory depth and width for a convolutional layer should be 
\begin{equation}
    M^{dep} = f_{t}c_{t}k_{t}^2,\;\; M^{wid} = f_{p}c_{p}k_{p}^2L_{W}
    \label{equ:dep_wid}
\end{equation}
respectively, to prevent any computation stalls within the PEs. The symbols used here are defined in Figure~\ref{fig:ce}.

One novelty of the proposed work is in the introduction of memory weight fragmentation. Under this scheme, the original weight memory structure is fragmented into static and dynamic regions (Figure~\ref{fig:fragment}), where the weights under the static regions are stored as in the conventional approaches, where the dynamic regions are sharing the same physical memory structure in a time-multiplexed manner.

Specifically, there are $n$ fragments stored in the on-chip memory, each with a depth of $u_{on}$; and $n$ fragments in the off-chip memory, each with a depth of $u_{o\!f\!f}$. The memory width remains the same as before. Therefore, the total depth of on-chip and off-chip memory can be represented as:
\begin{equation}
    M^{dep}_{on} = u_{on}n,\; M^{dep}_{o\!f\!f} = u_{o\!f\!f}n,\;  M^{dep} = M^{dep}_{on}+M^{dep}_{o\!f\!f}
    \label{equ:on_off_dep}
\end{equation}
The shared off-chip buffer is implemented with dual-port Block RAMs (BRAMs), supporting different clocks and port widths on the read and write sides, and allowing for independent control and data-transfer rates. As a result, the proposed architecture incorporates two distinct clock domains:

\begin{itemize}
    \item $clk_{comp}$: controls the execution of various computations within the CEs and the reading of the shared buffer.
    \item $clk_{dma}$: manages the process of loading weights from the off-chip memory and writing them into the shared buffer.
\end{itemize}

During run-time, the PE array iteratively reads weights between the static on-chip storage and the off-chip buffer, controlled by address counters and additional ``Read-After-Write" checking. This iterative process repeats $r$ times.
\begin{equation}
    r = b\hat{h}\hat{w}n,
    \label{equ:repeat}
\end{equation}
as convolution weights are reused on $b$, $\hat{h}$ and $\hat{w}$ dimensions (Figure~\ref{fig:ce}).

%\begin{algorithm}[t]
%    \caption{Memory Read Logic}
%    \label{algo:mem_read}
%    \footnotesize
%    \begin{algorithmic}
%        \Require $rptr_{pe}$ \Comment{$[0,f_{t}c_{t}k_{t}^{2})$, read pointer from PEs}
%        \State $i, j \gets \text{div}(rptr_{pe}, u_{on}\!+\!u_{o\!f\!f})$ \Comment{int division with remainder}
%        \If {$j\!<\!u_{on}$} 
%        \State $en_{on} \gets 1; en_{o\!f\!f} \gets 0$  \Comment{enable on-chip}
%        \State $rptr_{on} \gets u_{on}i+j$ \Comment{$[0,u_{on}n)$, read pointer for static on-chip storage}
%        \State \Return $en_{on}, en_{o\!f\!f}, rptr_{on}$
%        \Else 
%        \State $en_{on} \gets 0; en_{o\!f\!f} \gets 1$ \Comment{enable off-chip}
%        \State $rptr_{o\!f\!f} \gets j-u_{on}$ \Comment{$[0,u_{o\!f\!f})$, read pointer for off-chip buff}
%        \State \Return $en_{on}, en_{o\!f\!f}, rptr_{o\!f\!f}$
%        \EndIf
%    \end{algorithmic}
%\end{algorithm}
\begin{figure}[t]
    \centering
    \includegraphics[width=0.6\columnwidth]{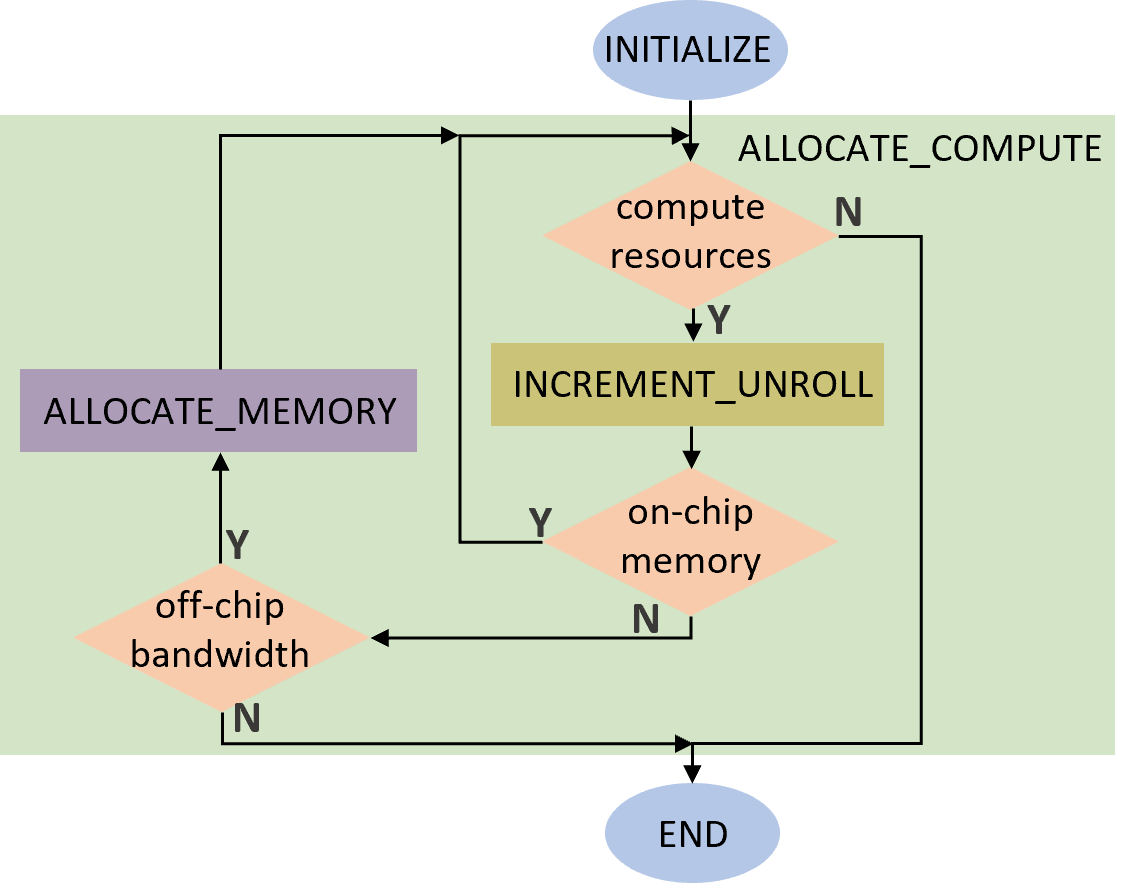}
    \caption{A high-level visualization of Algorithm~\ref{algo:dse}}
    \label{fig:algo_flow}
\end{figure}

\subsection{Resource and Performance Modelling}
In the proposed CE template, $k_p$, $c_p$ and $f_p$ control the parallelism of computations, while $n$, $u_{on}$ and $u_{o\!f\!f}$ dictate the memory storage structure (Figure~\ref{fig:ce}, Equation~\ref{equ:on_off_dep}). These variables, along with other user-defined parameters, including the computation clock frequency $clk_{comp}$, type of operation $O$, weights bitwidth $L_{W}$ and activations bitwidth $L_{A}$, define the configuration of the CE. The combination of these variables and parameters allows us to estimate the area $a$, the off-chip bandwidth $\beta$, and the throughput $\theta$ of a single CE.

\begin{align}
    \mathcal{V} = \{ k_p, c_p, f_p, n, u_{on}, u_{o\!f\!f} | clk_{comp}, O, L_{W}, L_{A} \}  \nonumber \\
    \Rightarrow a(\mathcal{V}), \beta(\mathcal{V}), \theta(\mathcal{V})
    \label{equ:abt}
\end{align}

The area estimation $a$ is calculated based on regression models developed by randomly sampling values of the tunable variables and collecting post-synthesis results. The throughput estimation $\theta$, is based on analytical models, leveraging the predictable parallelism behavior that can be analyzed in a cycle-accurate manner. Our approach to estimating area and throughput follows the same methodology as the previous work \cite{petrica2020memory, venieris2017fpgaconvnet}. In addition, specific to this work, The average off-chip bandwidth required by an individual CE is estimated by:
\begin{equation}
    \beta(\mathcal{V}) = M^{wid} \cdot clk_{comp} \cdot \frac{u_{o\!f\!f}}{u_{on}\!+\!u_{o\!f\!f}}
    \label{equ:bw_layer}
\end{equation}

Here, the product of the first two terms represents the number of bits required per second. The final scaling term accounts for the dual-port buffer's ability to load weights from off-chip memory, irrespective of whether the PEs are reading from on-chip storage or the off-chip buffer itself.

\section{Macro-architecture}
Let us denote the DNN model as $\mathcal{D}$, where each layer, denoted as $l \in \mathcal{D}$, is mapped to a CE on the hardware. To distinguish layer-specific values, we introduce the subscript $_l$ into previously defined symbols.

\subsection{Design Space Exploration}
Due to the layer-wise pipelined architecture, CEs are interconnected using FIFOs to accommodate variations in processing rates and data port width. Consequently, the overall throughput of the pipeline is determined by the slowest CE, leading to a resource-constrained optimization problem:

\begin{equation}
    \max(\min_{l \in \mathcal{D}}\theta_{l})\;\;\;\;s.t.\;\;\beta_{io} + \sum_{l \in \mathcal{D}} s_l\beta_{l}  \leq B, \;\; \sum_{l \in \mathcal{D}} a_{l}  \leq A
    \label{equ:dse}
\end{equation}

\begin{algorithm}[t]
    \caption{Greedy DSE}
    \label{algo:dse}
    \footnotesize
    \begin{algorithmic}
        \Require $\mathcal{D}$, $A$, $B$ \Comment{target DNN, area constraint, bandwidth constraint}
        \Procedure{initialize}{$\mathcal{D}$}
        \For{$l \in \mathcal{D}$}
            \State $k_{p\_l}, c_{p\_l}, f_{p\_l} \gets 1$  \Comment{minimize compute resources}
            \State $M^{dep}_{o\!f\!f\_l} \gets 0$ \Comment all weights on-chip
        \EndFor
        \EndProcedure
        \State

        \Procedure{delta\_bandwidth}{$\mathcal{D}, l$} 
            \State $\mathcal{D}' \gets \mathcal{D}; l' \gets l$; \Call{increment\_offchip}{$l'$}
            \State \Return $\Delta B \gets \sum_{l' \in \mathcal{D'}} s_{l'} \beta_{l'}$ - $\sum_{l \in \mathcal{D}} s_l\beta_{l}$ \Comment bandwidth difference
        \EndProcedure
        \Procedure{write\_burst\_balance}{$\mathcal{D}, l$} 
            \State $r_{max} = max(r_{l'},$ for $l' \in \mathcal{D}$ and $l'!=l)$ \Comment Equation~\ref{equ:balance}
            \State \Return $ r_{max}/(b\hat{h_l}\hat{w_l})$
        \EndProcedure
        \Procedure{increment\_offchip}{$\mathcal{D}, l$}
            \State $M^{dep}_{o\!f\!f\_l} \gets M^{dep}_{o\!f\!f\_l} + \mu$;   \Comment $\mu$, hyperparameter
            \State $n_l \gets$ \Call{write\_burst\_balance}{$\mathcal{D}, l$} 
        \EndProcedure
        \Procedure{allocate\_memory}{$\mathcal{D}$, $A$, $B$}
            \While {$\sum_{l \in \mathcal{D}} a_{l}^{mem} > A^{mem}$} \Comment{on-chip mem limit}
                \State $l \gets$ SORT\_BY($l \in \mathcal{D}$, \Call{delta\_bandwidth}{$\mathcal{D}, l$})[0]
                \State $\mathcal{D}' \gets \mathcal{D}; l' \gets l$; \Call{increment\_offchip}{$\mathcal{D}', l'$}
                \If {$\beta'_{io} +\sum_{l' \in \mathcal{D'}} s_{l'} \beta_{l'} > B$} \Return False \Comment{bandwidth limit}
                \EndIf
                \State $\mathcal{D} \gets \mathcal{D}'$
            \EndWhile
            \State \Return True
        \EndProcedure
        \State
        
        \Procedure{increment\_unroll}{$l$}
            \For{$v_l \in \{k^2_l, f_l, c_l\}$}
                \If{$v_{p\_l} < v_l$} 
                    \State $v_{p\_l} \gets v_{p\_l}+\phi$; \Return TRUE \Comment $\phi$, hyperparameter
                \EndIf
            \EndFor
             \State \Return FALSE
        \EndProcedure
        \Procedure{allocate\_compute}{$\mathcal{D}$, $A$, $B$}
            \While {$\sum_{l \in \mathcal{D}} a_{l} \leq A$} 
                \State $l \gets$ SORT\_BY($l \in \mathcal{D}$, $\theta_{l}$)[0] \Comment{slowest layer}
                \State $\mathcal{D}' \gets \mathcal{D}; l' \gets l$; $S_1 \gets $ \Call{increment\_unroll}{$l'$}
                \State $S_2 \gets$ \Call{allocate\_bandwidth}{$\mathcal{D}'$, $A^{mem}$, $B$}
                \If {$\sum_{l' \in \mathcal{D'}} a_{l'} > A'$ or !$S_1$ or !$S_2$} break \Comment{area limit}
                \EndIf
                \State $\mathcal{D} \gets \mathcal{D}'$
            \EndWhile
        \EndProcedure
        \State
        
        \State\Call{initialize}{$\mathcal{D}$}; \Call{allocate\_unroll}{$\mathcal{D}$, $A$, $B$}; 
        \State\Return $\min_{l \in \mathcal{D}}\theta_{l}$
    \end{algorithmic}
\end{algorithm}

$B$ and $A$ denote the device constraint on off-chip bandwidth and area respectively. $\beta_{io}$ accounts for the  bandwidth cost for the first PE to read input samples and the last PE to write prediction results, as illustrated in Figure~\ref{fig:arch_compare}. 

$s_{l}$ is defined as the ``slow-down" factor, which quantifies the throughput ratio between the slowest CE and the current CE. It accounts for situations where the processing rates of different CEs are not perfectly matched. In such cases, the required off-chip bandwidth can be scaled down proportionally, without impacting the overall pipeline throughput.

\begin{equation}
    s_l = \frac{\min_{l \in \mathcal{D}}\theta_{l}}{\theta_{l}}
\end{equation}

\begin{figure}[t]
    \centering
    \begin{minipage}{0.49\columnwidth}
        \includegraphics[trim={0.1cm 0 0.25cm 0},clip, width=\textwidth]{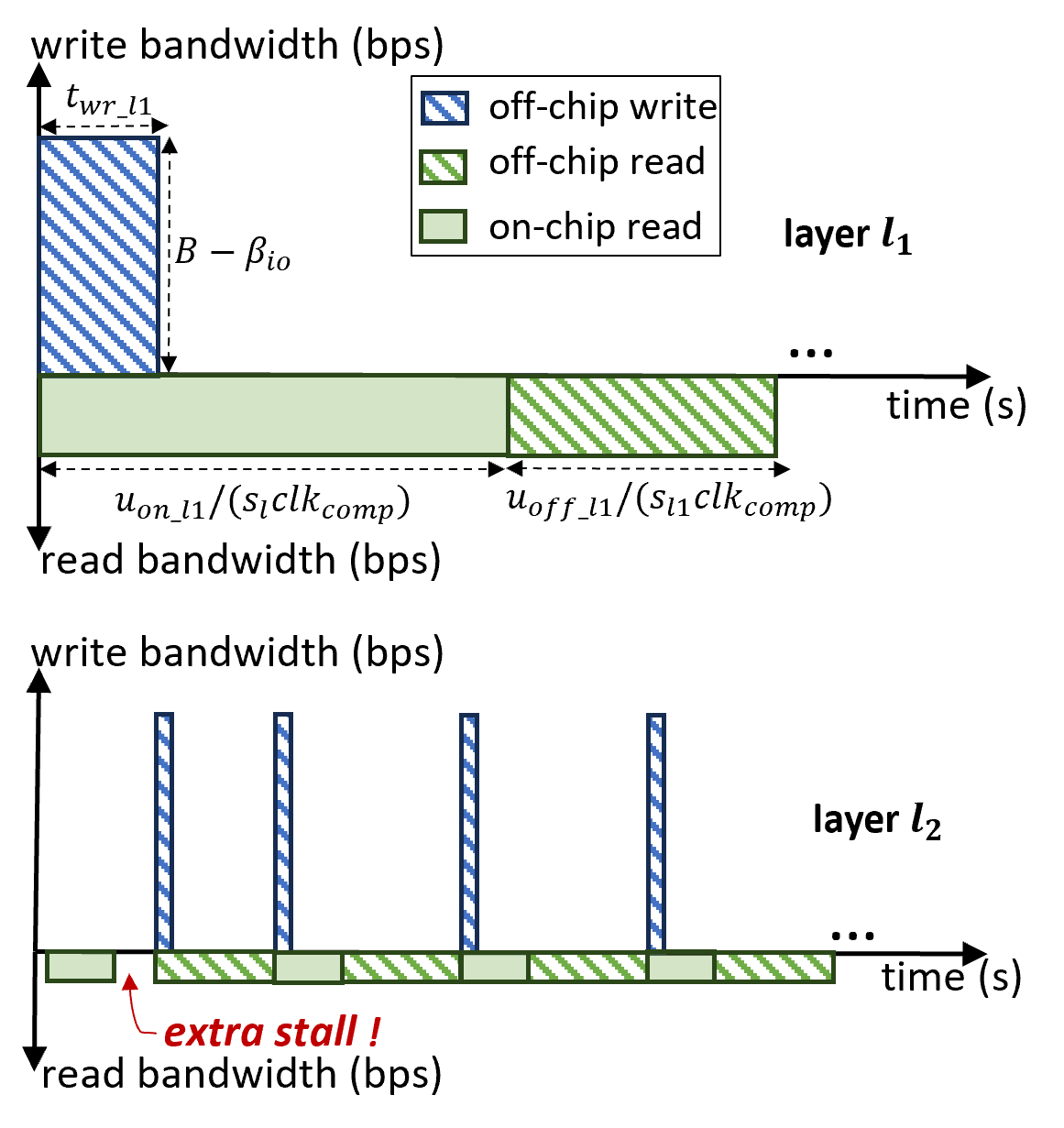}
        \subcaption{imbalanced burst numbers}
    \end{minipage}
    \hfill
    \begin{minipage}{0.49\columnwidth}
        \includegraphics[trim={0.1cm 0 0.3cm 0},clip, width=\textwidth]{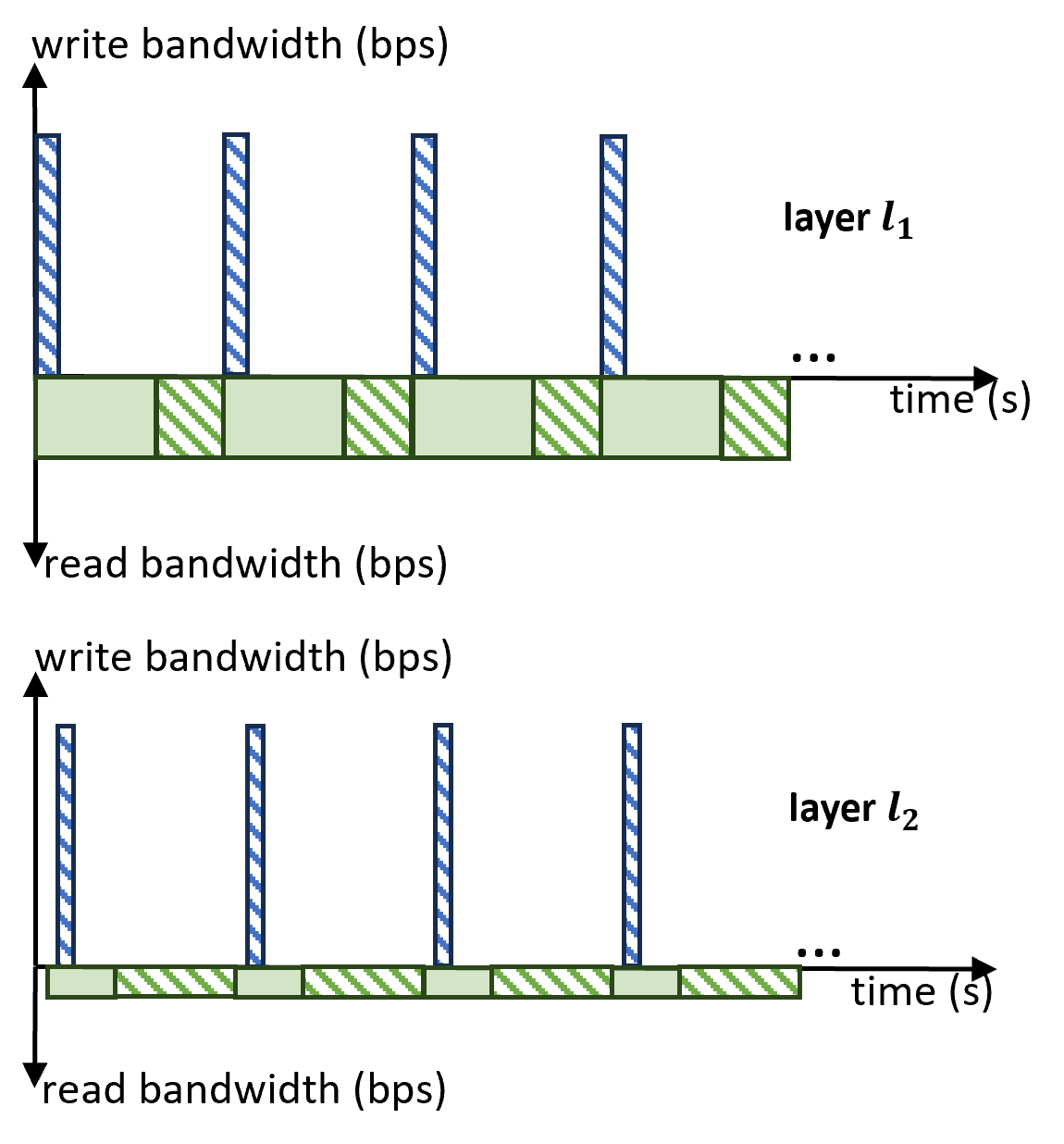}
        \subcaption{balanced burst numbers}
    \end{minipage}
    \caption{Two-layer example of write/read scheduling}
    \label{fig:schedule}
\end{figure}

The objective of Equation~\ref{equ:dse} is to maximize the accelerator's throughput by identifying the optimal combination of $\mathcal{V}$ (Equation~\ref{equ:abt}) for all CEs. However, conducting an exhaustive search can be time-consuming. To address this, our DSE method (Algorithm~\ref{algo:dse}) employs a greedy approach, which iteratively optimizes the computation and memory (Figure~\ref{fig:algo_flow}).

\begin{itemize}
    \item \textbf{greedy compute allocation}: In this step of the optimization, our heuristic is to incrementally promote the throughput of the slowest CE. We achieve this by iteratively increasing the unroll factors in dimensions such as $k_l^2, f_l, c_l$ by a user-defined step size $\phi$. After each adjustment, we re-evaluate the CE's throughput and repeat the process. When the allocated on-chip memory area exceeds the limit $A_{mem}$, we transition to the next phase, which focuses on off-chip bandwidth allocation.
    \item \textbf{greedy memory allocation}: Initially, we begin with a design where all weights reside on-chip. In each iteration, we select a layer and evict one memory block to off-chip memory. This block has a depth of $\mu$ and a width of $M^{wid}_l$ words. The choice of the layer is to minimize the marginal impact on bandwidth due to this eviction. Based on the total memory depth evicted to off-chip storage, denoted as $M^{dep}_{o\!f\!f\_l}$, the algorithm calculates the optimal number of memory fragments $n_l$ using a ``write burst balancing" strategy (explained in the following section). This calculation guides the determination of $u_{on\_l}$ and $u_{o\!f\!f\_l}$ as per Equation~\ref{equ:on_off_dep}.
\end{itemize}

In this DSE algorithm, two user-defined hyperparameters, $\phi$ and $\mu$, control the step sizes of the exploration. The choice of these hyperparameters affects the trade-off between exploration time and solution optimality. A larger step size accelerates exploration but may lead to sub-optimal solutions.

\subsection{DMA Connection and Scheduling}
\label{subsec:schedule}
At the microarchitecture level, we employ a demultiplexer to manage the routing between the DMA port and multiple CEs. The demultiplexer is controlled by a configuration sequence that outlines the order and the duration of serving each individual CE. 

In the \textbf{top-left} region of Figure~\ref{fig:schedule}, we present the write and read scheduling for layer $l_{1}$. Writing operates in burst mode to fill the off-chip memory buffer, fully utilizing the available device bandwidth, $B\!-\!\beta_{io}$, after deducting the input and output transmissions. Therefore, we can calculate the duration of this write burst as:
\begin{equation}
    t_{wr\_l1} = \frac{M^{wid}_{l1} \cdot u_{o\!f\!f\_l1}}{(B\!-\!\beta_{io})}
\end{equation}
The interval between burst writes is the sum of time spent on reading the static on-chip storage and off-chip buffer:
\begin{equation}
    t_{rd\_l1} = \frac{u_{on\_l1}+u_{o\!f\!f\_l1}}{s_{l1}clk_{comp}}
\end{equation}
And this pattern will repeat for $r_{l1}$ times (Equation~\ref{equ:repeat}). 

Moving to the \textbf{bottom-left} of the figure, which displays the scheduling of the layer $l_{2}$ instead. Here, we assume layer $l_{2}$ is connected after $l_{1}$ in the pipeline. Consequently, the first read of layer $l_{2}$ begins slightly later than that of layer $l_{1}$ due to the pipeline depth between these two layers. In addition, $r_{l2}$ is four times $r_{l1}$, so as the number of burst writes. However, this mismatch in the number of bursts introduces additional stalls in $l_{2}$ when the DMA is occupied with writing the substantial weight chunk for $l_{1}$. These stalls occur $r_{l1}$ times in total, affecting overall performance.

In constrast, looking at the \textbf{top-right} and the \textbf{bottom-right} of Figure~\ref{fig:schedule}, where $r_{l1}$ is set equal as $r_{l2}$, those stalls are diminished. Therefore, we employ the ``write burst balancing" strategy in Algorithm~\ref{algo:dse}, that enforces the following condition:
\begin{equation}
    \forall l_1, l_2 \in D,\;r_{l1} = r_{l2} 
    \label{equ:balance}
\end{equation}
This condition ensures that the number of bursts for different layers is equal, avoiding the stalls and optimizing overall performance. 

\begin{table}[t]
    \caption{Characteristics of evaluated models. The accuracy of some quantized models is higher than the floating point versions due to extra fine-tuning.}
    \centering
    \scriptsize
    \setlength{\tabcolsep}{3pt} 
    \begin{tabular}{lllllll}
        \toprule
         \multirow{2}{*}{Network}   & \multicolumn{4}{c}{ImageNet Accuracy} & \multirow{2}{*}{Params} & \multirow{2}{*}{MACs} \\
         \cmidrule{2-5}   
         ~ & W4A4\cite{chang2021mix} & W4A5\cite{sun2022film} & W8A8\cite{kathail2020xilinx} & F32 & ~ & ~  \\
         \midrule
        mobilenetv2 & 65.6 & 65.7 & 67.7 & 71.9 & 3.5M & 0.3G \\
        resnet18 & 70.3 & 70.5 & 70.0 & 69.8 & 11.7M & 1.8G \\
        resnet50 & - & 77.3 & 76.0 & 76.1 & 25.6M & 4.1G \\
        \bottomrule
    \end{tabular}
    \label{tab:network}
\end{table}
\begin{table*}
    \caption{Latency (ms) results across different networks and devices. $^*$ denotes W4A4, $^\dag$ denotes W4A5 , $^\diamond$ denotes W8A8}
    \centering
    \scriptsize
    \setlength{\tabcolsep}{2pt}    
    \begin{minipage}{0.32\textwidth}
    \begin{tabular}{llll}
    \toprule
       \multicolumn{4}{c}{mobilenetv2} \\
    \midrule
       \multirow{2}{*}{Architecture} & \multicolumn{3}{c}{Device} \\
    \cmidrule(lr){2-4}
       ~ & Zedboard & ZC706 & ZCU102  \\
    \midrule
        layer-sequential & \textcolor{ForestGreen}{8.3}$^*$\cite{chang2021mix} & 7.3$^*$\cite{chang2021mix} & 5.3$^\dag$\cite{sun2022film}\\ % chang2021mix results from https://dl.acm.org/doi/pdf/10.1145/3490422.3502367
    \midrule
        vanilla layer-pipelined & \xmark & 9.2$^*$ & 2.3$^\dag$\\
    \midrule 
        this work & 325.9$^*$ & \textcolor{ForestGreen}{4.8}$^*$ & \textcolor{ForestGreen}{2.3}$^\dag$ \\
    \bottomrule   
    \end{tabular}
    \end{minipage}
    \begin{minipage}{0.32\textwidth}
    \begin{tabular}{llll}
    \toprule
       \multicolumn{4}{c}{resnet18} \\
    \midrule
       \multirow{2}{*}{Architecture} & \multicolumn{3}{c}{Device} \\
    \cmidrule(lr){2-4}
       ~ & ZC706 & ZCU102 & U50 \\
    \midrule
        layer-sequential & 40.4$^*$\cite{chang2021mix} & 13.7$^\dag$\cite{sun2022film} & 3.0$^\diamond$\cite{kathail2020xilinx}  \\ % https://docs.xilinx.com/r/1.1-English/ug1354-xilinx-ai-sdk/ZCU104-Performance, chen2021bring
    \midrule
        vanilla layer-pipelined & \xmark & \xmark & 1.3$^\diamond$ \\
    \midrule 
        this work & \textcolor{ForestGreen}{27.0}$^*$ & \textcolor{ForestGreen}{7.0}$^\dag$ & \textcolor{ForestGreen}{1.3}$^\diamond$ \\
    \bottomrule   
    \end{tabular}
    \end{minipage}
    \begin{minipage}{0.32\textwidth}
    \begin{tabular}{llll}
    \toprule
       \multicolumn{4}{c}{resnet50} \\
    \midrule
       \multirow{2}{*}{Architecture} & \multicolumn{3}{c}{Device} \\
    \cmidrule(lr){2-4}
       ~ & ZCU102 & U50 & U250 \\
    \midrule
        layer-sequential & \textcolor{ForestGreen}{21.1}$^\dag$\cite{sun2022film} & 6.0$^\diamond$\cite{kathail2020xilinx} & 5.6$^\diamond$\cite{kathail2020xilinx} \\
    \midrule
        vanilla layer-pipelined & \xmark & 15.0$^\diamond$ & 1.8$^\diamond$  \\
    \midrule 
        this work & 578.7$^\dag$ & \textcolor{ForestGreen}{3.4$^\diamond$} & \textcolor{ForestGreen}{1.8$^\diamond$} \\
    \bottomrule   
    \end{tabular}
    \end{minipage}
    \label{tab:overall_compare}
\end{table*}

\section{Evaluation}
\subsection{Experiment Setup}

We target the AMD Xilinx FPGA devices and use Vivado 2019.1 for hardware synthesis. Regarding the DNNs, we deploy the quantized models provided by existing research \cite{chang2021mix, sun2022film, kathail2020xilinx}, and their accuracy, number of Multiply–ACcumulate (MAC) operation, and parameters are summarized in Table~\ref{tab:network}.

We extend fpgaConvNet \cite{venieris2017fpgaconvnet}, which is an open-source toolflow, that generates layer-wise pipelined accelerators. We build our own weights fragmentation (Figure~\ref{fig:fragment}), DSE method (Algorithm~\ref{algo:dse}), and DMA scheduling (Figure~\ref{fig:schedule}) upon that toolflow. In the rest of the paper, we refer to the original fpgaConvNet, which did not exploit off-chip weights storage, as the ``vanilla layer-pipelined" approach.

As our design methodology is fully automated, it can also be easily integrated into other layer-wise pipelined toolflows, such as FINN \cite{petrica2020memory} and hls4ml \cite{fahim2021hls4ml}, in the future.

\subsection{Overall Results}
Table~\ref{tab:overall_compare} provides a comparative analysis of our methodology, the ``vanilla layer-pipelined" approach, and other ``layer-sequential" architectures. We define device size relative to model parameters; for example, ZCU102 is ``large" for MobileNetV2 but ``small" for ResNet50 due to its larger parameter size. Key observations include:

\begin{itemize}
    \item ``Vanilla layer-pipelined" excels on ``large" FPGA devices with ample on-chip memory. For example, mapping MobileNetV2 to ZCU102 achieves 2.3ms latency, less than half of ``layer-sequential" (5.3ms). Similar trends apply to ResNet18 on U50 and ResNet50 on U250.
    \item Our methodology maintains latency on these ``large" devices, as our greedy DSE automatically determines that there is no need to store weights on-chip, and the designs become primarily compute-bound.
    \item When on-chip memory resources become bottleneck, the ``vanilla layer-pipelined" approach may become inferior to ``layer-sequential" or, in some cases, may not fit the device at all (marked as ``\xmark" in the table).
    \item The advantage of our proposed methodology becomes evident on these ``smaller" devices. For example when mapping ResNet50 to U50, our approach reduces the latency from 15.0ms (in the ``vanilla layer-pipelined" approach) to 3.4ms. It also surpasses the ``layer-sequential" approach, which requires 6.0ms.
    \item In some cases, such as MobileNetV2 on Zedboard and ResNet50 on ZCU102, ``layer-sequential" achieves the lowest latency. This is because the off-chip bandwidth on these devices is limited compared to the number of parameters in those models. This bandwidth constraint restricts the full application of our proposed methodology.  Additionally, the overhead of implementing per-layer FIFOs and buffers becomes significant in these cases.
\end{itemize}

In summary, the results in Table~\ref{tab:overall_compare} demonstrate that ``layer-pipelined" approaches have a clear advantage over ``layer-sequential" approaches on ``large" devices with ample on-chip memory resources. Our work extends this advantage to more resource-constrained devices by considering the access requirements of the weights and leveraging off-chip memory bandwidth.
\begin{figure}[t]
    \centering
    \begin{subfigure}[b]{0.48\columnwidth}
    \centering
        \begin{tikzpicture}[thick,scale=0.6, every node/.style={scale=0.82}]

\pgfplotsset{
    tick label style={font=\large}, % Adjust the font size for tick labels
    label style={font=\large}, % Adjust the font size for axis labels
}
    \begin{axis}[
    width=8cm,
    height=6cm,
    xlabel={Normalized On-chip Memory ($A^{mem}$)},
    ylabel={FPS},
    ylabel style={yshift=-8pt},
]

\addplot[orange!70, only marks, mark=*, mark size=2] coordinates {
(1.25,1.728353519364607)
(1.3,	6.894692534633764)
(1.35,13.738970011400596)
(1.4, 6.895422301519578)
(1.45, 13.748395562237887)
(1.5, 6.896973866503898)
(1.55, 13.748031711484908)
(1.6, 1.7285140069657212)
(1.65, 1.7285712842357264)
(1.7, 10.33117298175112)
(1.72, 140.7074347697921)
(1.75, 269.7788353108122)
(1.8, 269.7788353108122)
(1.85, 269.7788353108122)
(1.9, 269.7788353108122)
(1.95, 269.7788353108122)
(2, 269.7788353108122)
};

\addplot[draw=orange!70, dotted, line width=1pt] coordinates {
(1.25,1.728353519364607)
(1.3,	6.894692534633764)
(1.35,13.738970011400596)
(1.4, 6.895422301519578)
(1.45, 13.748395562237887)
(1.5, 6.896973866503898)
(1.55, 13.748031711484908)
(1.6, 1.7285140069657212)
(1.65, 1.7285712842357264)
(1.7, 10.33117298175112)
(1.72, 140.7074347697921)
(1.75, 269.7788353108122)
(1.8, 269.7788353108122)
(1.85, 269.7788353108122)
(1.9, 269.7788353108122)
(1.95, 269.7788353108122)
(2, 269.7788353108122)
};

\addplot[blue!70, only marks, mark=x, mark size=3] coordinates {
(0.05,	12.62629849642881)
(0.1,	14.426347473149502)
(0.15,	16.959704928269776)
(0.2,	22.718551901326965)
(0.25,	23.612427503944456)
(0.3,	28.677736942273576)
(0.35,	34.898843085902286)
(0.4,	37.10186911796243)
(0.45,	47.72640931314061)
(0.5,	53.070837893002825)
(0.55,	60.56133697629963)
(0.6,	73.08990496850556)
(0.65,	73.08990496850556)
(0.7,	 87.07740789713719)
(0.75,	94.53229908695978)
(0.8,	107.11041798769516)
(0.85,	110.8330433202033)
(0.9,	126.52077977286989)
(0.95,	142.22920093501477)
(1,	141.98374002209266)
(1.05,	148.0266566403278)
(1.1,	191.69812912210884)
(1.15,	199.9558097660417)
(1.2,	212.3331592201428)
(1.25,	212.3331592201428)
(1.3,	226.9498395464634)
(1.35,	271.79600079364434)
(1.4,	269.7788353108122)
(1.45,	269.7788353108122)
(1.5,	269.7788353108122)
(1.55,	269.7788353108122)
(1.6,	269.7788353108122)
(1.65,	269.7788353108122)
(1.7,	269.7788353108122)
(1.75,	269.7788353108122)
(1.8,	269.7788353108122)
(1.85,	269.7788353108122)
(1.9,	269.7788353108122)
(1.95,	269.7788353108122)
(2,	269.7788353108122)
};

\addplot[draw=blue!70,  dotted, line width=1pt] coordinates {
(0.05,	12.62629849642881)
(0.1,	14.426347473149502)
(0.15,	16.959704928269776)
(0.2,	22.718551901326965)
(0.25,	23.612427503944456)
(0.3,	28.677736942273576)
(0.35,	34.898843085902286)
(0.4,	37.10186911796243)
(0.45,	47.72640931314061)
(0.5,	53.070837893002825)
(0.55,	60.56133697629963)
(0.6,	73.08990496850556)
(0.65,	73.08990496850556)
(0.7,	 87.07740789713719)
(0.75,	94.53229908695978)
(0.8,	107.11041798769516)
(0.85,	110.8330433202033)
(0.9,	126.52077977286989)
(0.95,	142.22920093501477)
(1,	141.98374002209266)
(1.05,	148.0266566403278)
(1.1,	191.69812912210884)
(1.15,	199.9558097660417)
(1.2,	212.3331592201428)
(1.25,	212.3331592201428)
(1.3,	226.9498395464634)
(1.35,	271.79600079364434)
(1.4,	269.7788353108122)
(1.45,	269.7788353108122)
(1.5,	269.7788353108122)
(1.55,	269.7788353108122)
(1.6,	269.7788353108122)
(1.65,	269.7788353108122)
(1.7,	269.7788353108122)
(1.75,	269.7788353108122)
(1.8,	269.7788353108122)
(1.85,	269.7788353108122)
(1.9,	269.7788353108122)
(1.95,	269.7788353108122)
(2,	269.7788353108122)
};

\draw[->] (80,141.98374002209266+25) -- (92,141.98374002209266+8);
\node[anchor=west] at (70,141.98374002209266+37) {$d_1$};

\draw[->] (180,140.7074347697921+25) -- (170,140.7074347697921+5);
\node[anchor=west] at (175,140.7074347697921+37) {$d_0$};

\end{axis}
\end{tikzpicture}
    \end{subfigure}
    \hfill
    \begin{subfigure}[b]{0.48\columnwidth}
    \centering
        \begin{tikzpicture}[thick,scale=0.6, every node/.style={scale=0.82}]

\pgfplotsset{
    tick label style={font=\large}, % Adjust the font size for tick labels
    label style={font=\large}, % Adjust the font size for axis labels
}

\begin{axis}[
    width=8cm,
    height=6cm,
    xlabel={Normalized On-chip Memory ($A^{mem}$)},
    ylabel={Normalized Off-chip Bandwidth},
    ylabel style={yshift=-8pt}
]

\addplot[orange!70, only marks, mark=*, mark size=2] coordinates {
(1.25,0.0013107334405116212/153.6)
(1.3,	0.005242933762046485/153.6)
(1.35,0.01048586752409297/153.6)
(1.4, 0.005242933762046485/153.6)
(1.45, 0.01048586752409297/153.6)
(1.5, 0.005242933762046485/153.6)
(1.55, 0.01048586752409297/153.6)
(1.6, 0.0013107334405116212/153.6)
(1.65, 0.0013107334405116212/153.6)
(1.7, 0.007864400643069728/153.6)
(1.72, 0.2283588927469136/153.6)
(1.75, 0.2283588927469136/153.6)
(1.8, 0.2283588927469136/153.6)
(1.85, 0.2283588927469136/153.6)
(1.9, 0.2283588927469136/153.6)
(1.95, 0.2283588927469136/153.6)
(2, 0.2283588927469136/153.6)
};

\addlegendentry{Vanilla}

\addplot[draw=orange!70, dotted, line width=1pt, forget plot] coordinates {
(1.25,0.0013107334405116212/153.6)
(1.3,	0.005242933762046485/153.6)
(1.35,0.01048586752409297/153.6)
(1.4, 0.005242933762046485/153.6)
(1.45, 0.01048586752409297/153.6)
(1.5, 0.005242933762046485/153.6)
(1.55, 0.01048586752409297/153.6)
(1.6, 0.0013107334405116212/153.6)
(1.65, 0.0013107334405116212/153.6)
(1.7, 0.007864400643069728/153.6)
(1.72, 0.2283588927469136/153.6)
(1.75, 0.2283588927469136/153.6)
(1.8, 0.2283588927469136/153.6)
(1.85, 0.2283588927469136/153.6)
(1.9, 0.2283588927469136/153.6)
(1.95, 0.2283588927469136/153.6)
(2, 0.2283588927469136/153.6)
};

\addplot[blue!70, only marks, mark=x, mark size=2] coordinates {
(0.05,	152.25677951663062/153.6)
(0.1,	149.5045643982356/153.6)
(0.15,	152.20729164998502/153.6)
(0.2,	152.12286413970872/153.6)
(0.25,	146.86960306582301/153.6)
(0.3,	147.58029933433488/153.6)
(0.35,	152.1567988521342/153.6)
(0.4,	137.266130791565/153.6)
(0.45,	152.22214224531933/153.6)
(0.5,	152.6433857764014/153.6)
(0.55,	152.33447746635315/153.6)
(0.6,	151.15741517885073/153.6)
(0.65,	135.05913508057063/153.6)
(0.7,	 152.32529586257613/153.6)
(0.75,	152.3010236880986/153.6)
(0.8,	152.1188135493906/153.6)
(0.85,	135.3371024120744/153.6)
(0.9,	152.24080517343992/153.6)
(0.95,	135.03776663773152/153.6)
(1,	105.0997631103946/153.6)
(1.05,	129.4778452975792/153.6)
(1.1,	152.1309587879312/153.6)
(1.15,	129.96343635149572/153.6)
(1.2,	139.0811276646025/153.6)
(1.25,	139.0811276646025/153.6)
(1.3,	141.31579722453884/153.6)
(1.35,	131.1762742104539/153.6)
(1.4,	109.53466824697293/153.6)
(1.45,	52.141751382458835/153.6)
(1.5,	0.886120479470249/153.6)
(1.55,	0.2283588927469136/153.6)
(1.6,	0.2283588927469136/153.6)
(1.65,	0.2283588927469136/153.6)
(1.7,	0.2283588927469136/153.6)
(1.75,	0.2283588927469136/153.6)
(1.8,	0.2283588927469136/153.6)
(1.85,	0.2283588927469136/153.6)
(1.9,	0.2283588927469136/153.6)
(1.95,	0.2283588927469136/153.6)
(2,	0.2283588927469136/153.6)
};

\addlegendentry{AutoWS}

\addplot[draw=blue!70,  dotted, line width=1pt, forget plot] coordinates {
(0.05,	152.25677951663062/153.6)
(0.1,	149.5045643982356/153.6)
(0.15,	152.20729164998502/153.6)
(0.2,	152.12286413970872/153.6)
(0.25,	146.86960306582301/153.6)
(0.3,	147.58029933433488/153.6)
(0.35,	152.1567988521342/153.6)
(0.4,	137.266130791565/153.6)
(0.45,	152.22214224531933/153.6)
(0.5,	152.6433857764014/153.6)
(0.55,	152.33447746635315/153.6)
(0.6,	151.15741517885073/153.6)
(0.65,	135.05913508057063/153.6)
(0.7,	 152.32529586257613/153.6)
(0.75,	152.3010236880986/153.6)
(0.8,	152.1188135493906/153.6)
(0.85,	135.3371024120744/153.6)
(0.9,	152.24080517343992/153.6)
(0.95,	135.03776663773152/153.6)
(1,	105.0997631103946/153.6)
(1.05,	129.4778452975792/153.6)
(1.1,	152.1309587879312/153.6)
(1.15,	129.96343635149572/153.6)
(1.2,	139.0811276646025/153.6)
(1.25,	139.0811276646025/153.6)
(1.3,	141.31579722453884/153.6)
(1.35,	131.1762742104539/153.6)
(1.4,	109.53466824697293/153.6)
(1.45,	52.141751382458835/153.6)
(1.5,	0.886120479470249/153.6)
(1.55,	0.2283588927469136/153.6)
(1.6,	0.2283588927469136/153.6)
(1.65,	0.2283588927469136/153.6)
(1.7,	0.2283588927469136/153.6)
(1.75,	0.2283588927469136/153.6)
(1.8,	0.2283588927469136/153.6)
(1.85,	0.2283588927469136/153.6)
(1.9,	0.2283588927469136/153.6)
(1.95,	0.2283588927469136/153.6)
(2,	0.2283588927469136/153.6)
};

\draw[->] (80,680) -- (92,680);
\node[anchor=west] at (65,680) {$d_1$};

\draw[->] (172,120) -- (172,30);
\node[anchor=west] at (165,170) {$d_0$};

\end{axis}
\end{tikzpicture}
    \end{subfigure}
    \vspace*{-5mm}
    \caption{resnet18-ZCU102, memory and performance trade-off. Normalization is against the max resource available on device}
    \label{fig:resnet18_fps_bw}
\end{figure}
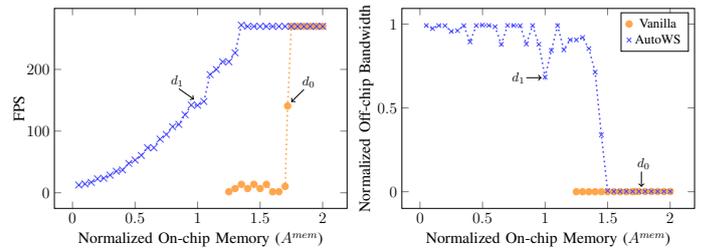

\begin{table}[t]
    \caption{resnet18-ZCU102, memory resource breakdown}
    \centering
    \scriptsize
    \setlength{\tabcolsep}{1.5pt}
    \begin{tabular}{llllllllll}
    \toprule
      \multirow{2}{*}{Design Point} & \multicolumn{3}{c}{Off-chip BW (Gbps)} & \multicolumn{4}{c}{BRAM Usage (MB)} & \multirow{2}{*}{DSP} & \multirow{2}{*}{FPS} \\
    \cmidrule(lr){2-4}\cmidrule(lr){5-8}
       & act & wt & total (util.) & act\_fifo & wt\_buff & wt\_mem & total(util.) & ~ & ~ \\
    \midrule
       Vanilla ($d_0$) & 0.1 & 0.0 & 0.1 (0\%) & 0.4 & 0.0 & 8.3 & 8.7 (\textcolor{red}{172\%}) & 1113 & 141 \\ % baseline 1.72
       AutoWS ($d_1$) & 0.1 & 105.0 & 105.1 (68\%) & 0.4 & 0.1 & 4.6 & 5.1 (99\%) & 1180 & 142 \\  % stream 1
    \bottomrule
    \end{tabular}
    \label{tab:mem_breakdown}
\end{table}

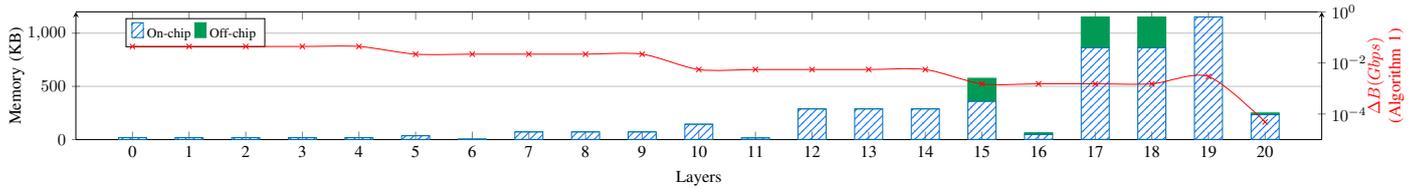
\begin{figure*}
    \centering
    \begin{tikzpicture}[thick,scale=0.7, every node/.style={scale=0.9}]
\begin{axis}[
    width=25cm,
    height=4cm,
    axis y line=left,
    ybar stacked,
    bar width=15pt,
    xlabel={Layers},
    ylabel={Memory (KB)},
    xtick=data,
    xticklabels={0, 1, 2, 3, 4, 5, 6, 7, 8, 9, 10, 11, 12, 13, 14, 15, 16, 17, 18, 19, 20, 21}, % Add your layer names here
   legend style={
    at={(0.04,0.95)}, % Adjust the coordinates as needed
    anchor=north west, % Position the legend in the top left
    legend columns=-1,
    font=\footnotesize,
    },
    xmin=0,
    xmax=22,
    ymin=0,
    ymax=1200, % Adjust as needed
    ymajorgrids=true
]
% On-chip memory data
\addplot[ybar, draw=NavyBlue, pattern=north east lines, pattern color=NavyBlue] coordinates
    {(1, 18.375) (2, 18) (3, 18) (4, 18) (5, 18) (6, 36) (7, 4) (8, 72) (9, 72) (10, 72) (11, 144) (12, 16) (13, 288) (14, 288) (15, 288) (16, 360) (17, 48) (18, 864) (19, 864) (20, 1152) (21, 234)};
% Off-chip memory data
\addplot[ybar, draw=ForestGreen, fill=ForestGreen] coordinates
    {(1, 0) (2, 0) (3, 0) (4, 0) (5, 0) (6, 0) (7, 0) (8, 0) (9, 0) (10, 0) (11, 0) (12, 0) (13, 0) (14, 0) (15, 0) (16, 216) (17, 16) (18, 288) (19, 288) (20, 0) (21, 16)};

\legend{On-chip, Off-chip}

\end{axis}

\begin{axis}[
    width=25cm,
    height=4cm,
    axis y line=right,
    ylabel style={yshift=-5pt,align=center},
    ylabel=\textcolor{red}{$\Delta B (Gbps)$} \\ \textcolor{red}{(Algorithm~\ref{algo:dse})},   % Right y-axis label
    xmin=0,
    xmax=22,
    ymin=1/100000,                      % Minimum value for the right y-axis
    ymax=1,                    % Maximum value for the right y-axis
    hide x axis,                % Hide the x-axis (shared with the left)
    ymode=log,
]
\addplot[smooth,color=red,mark=x]
    plot coordinates {
        (1,462848/512*8*0.2/32768)
        (2,462848/512*8*0.2/32768)
        (3,462848/512*8*0.2/32768)
        (4,462848/512*8*0.2/32768)
        (5,462848/512*8*0.2/32768)
        (6, 462848/1024*8*0.2/32768)
        (7, 462848/1024*8*0.2/32768)
        (8, 462848/1024*8*0.2/32768)
        (9, 462848/1024*8*0.2/32768)
        (10, 462848/1024*8*0.2/32768)
        (11, 462848/4096*8*0.2/32768)
        (12, 462848/4096*8*0.2/32768)
        (13, 462848/4096*8*0.2/32768)
        (14, 462848/4096*8*0.2/32768)
        (15, 462848/4096*8*0.2/32768)
        (16, 462848/14848*8*0.2/32768)
        (17, 462848/14848*8*0.2/32768)
        (18, 462848/14848*8*0.2/32768)
        (19, 462848/14848*8*0.2/32768)
        (20, 462848/7680*8*0.2/32768)
        (21, 462848/462848*8*0.2/32768)
    };
\end{axis}
\end{tikzpicture}
    \vspace*{-5mm}
    \caption{resnet18-ZCU102, per-layer on-chip and off-chip allocation for the design point $d_1$}
    \label{fig:resnet18_ratio}
\end{figure*}

\subsection{Case Study: resnet18-ZCU102}
In this section, we provide a case study that focuses on mapping ResNet18 to ZCU102, offering detailed insights into our implementation. Firstly, we conducted a parameter sweep, as depicted in Figure~\ref{fig:resnet18_fps_bw}, systematically adjusting the budget of on-chip memory ($A^{mem}$),  while keeping the budgets of compute resource (LUT, DSP) and off-chip bandwidth fixed. All resource numbers are normalized to the specifications of a single ZCU102 device.

Based on the value of $A^{mem}$, Figure~\ref{fig:resnet18_fps_bw} on the left can be split into three regions:

\begin{itemize}
    \item $[0, 1.25)$: Here, the ``vanilla" approach cannot fit, resulting in no provided design points. However, AutoWS exhibits steadily improved throughput as on-chip memory resources increase.
    \item $[1.25, 1.75)$: The ``vanilla" approach is feasible but lags behind AutoWS in throughput, suggesting the bottleneck changes from the memory capacity to the bandwidth.
    \item $[1.75, 2)$: In this range, both the ``vanilla" approach and AutoWS converge to the same set of design points as the accelerator becomes compute-bound.
\end{itemize}

Table~\ref{tab:mem_breakdown} offers a detailed resource breakdown for two specific design points, namely $d_0$ and $d_1$, selected from Figure~\ref{fig:resnet18_fps_bw}.  These design points correspond to the ``vanilla" approach and AutoWS, respectively. For $d_0$, its off-chip bandwidth is notably underutilized, as the ``vanilla" approach did not account for weight transfers (wt). Conversely, AutoWS ($d_1$) effectively utilizes this resource.

Regarding BRAM usage, it is calculated as the product between the number of BRAMs and the maximum capacity per BRAM. The usage falls into three categories:
\begin{itemize}
 \item \textbf{act\_fifo}: FIFOs and buffers connecting PEs and storing intermediate activation data.
 \item \textbf{wt\_buff}: buffers for loading off-chip weights.
 \item \textbf{wt\_mem}: static on-chip weight storage (Figure~\ref{fig:fragment}).
\end{itemize}
The costs of \textbf{act\_fifo} and \textbf{wt\_buff} are relatively minor (below 10\%) compared to \textbf{wt\_mem}. A comparison between design points $d_0$ and $d_1$, with similar throughput, reveals that AutoWS saves BRAM utilization by 70\%.

Furthermore, Figure~\ref{fig:resnet18_ratio} shows the layer-wise memory allocation of our DSE algorithm. In this case, 5 out of 21 layers have part of the weights stored off-chip (layers 15 to 18 and 20). The selection of these layers prioritizes minimal bandwidth impact with smaller $\Delta B$, as outlined in Algorithm~\ref{algo:dse}. We visualize this criterion as the red curve in Figure~\ref{fig:resnet18_ratio}, with the corresponding values obtained at the end of DSE. The savings in BRAM in Table~\ref{tab:mem_breakdown} are actually larger than the size of the off-chip weights in Figure~\ref{fig:resnet18_ratio}, as they are two different metrics. Some BRAMs were not fully filled in $d_0$ and the corresponding weights are now moved off the chip in $d_1$.

\subsection{Object Detection}
Furthermore, we evaluate the effectiveness of our proposed methodology in the context of the COCO object detection task, using the quantized to 8 bits YOLOv5n model, and targeting the ZCU102. AutoWS (8.7ms) achieves a $36\%$ latency reduction compared to Vitis AI (13.7ms) \cite{kathail2020xilinx}, and a $9\%$ reduction compared to the ``vanilla layer-pipelined" (9.5ms).

%\begin{table}[h]
%    \caption{Latency comparison on object detection task}
%    \centering
%    \scriptsize
%    \setlength{\tabcolsep}{4pt}
%    \begin{tabular}{llll}
%    \toprule
%       \multicolumn{4}{c}{YOLOv5n-ZCU102, W8A8} \\
%    \midrule
%       ~ & layer-sequential & vanilla layer-pipelined & this work \\
%    \midrule
%       Latency  & 13.7ms \cite{kathail2020xilinx} (1.00$\times$) & 9.5ms (0.69$\times$)  & \textcolor{ForestGreen}{8.7ms (0.64$\times$)} \\
%    \bottomrule   
%    \end{tabular}
%    \label{tab:yolo}
%\end{table}

\section{Conclusion}
In this paper, we introduced AutoWS, a novel memory management methodology capable of partially reloading weights from off-chip memory and efficiently delivering them to multiple pipelined CEs. Our hardware design, which is template-based and adaptable through a greedy DSE process, builds upon the advantages of ``vanilla layer-pipelined" approaches over ``layer-sequential" architectures. Moreover, we extend these benefits to resource-constrained devices. Future work would explore software-hardware co-design, such as weight encoding and pruning methods, to further enhance performance.

\section*{Acknowledgement}
For the purpose of open access, the author(s) has applied a Creative Commons Attribution (CC BY) license to any Accepted Manuscript version arising.

\bibliographystyle{IEEEtran}
\bibliography{bibliography}

\end{document}